# Field-effect tunneling transistor based on vertical graphene heterostructures

L. Britnell[1], R. V. Gorbachev[2], R. Jalil[2], B. D. Belle[2], F. Schedin[2], M. I. Katsnelson[3], L. Eaves[4], S. V. Morozov[5], N. M. R. Peres[6], J. Leist[7], A. K. Geim[1,2], K. S. Novoselov[1], L. A. Ponomarenko[1*]

[1]School of Physics & Astronomy, University of Manchester, Manchester M13 9PL, UK
[2]Manchester Centre for Mesoscience & Nanotechnology, University of Manchester, Manchester M13 9PL, UK
[3]Institute for Molecules and Materials, Radboud University of Nijmegen, 6525 AJ Nijmegen, The Netherlands
[4]School of Physics & Astronomy, University of Nottingham, Nottingham NG7 2RD, UK
[5]Institute for Microelectronics Technology, 142432 Chernogolovka, Russia
[6]Departamento de Física, Universidade do Minho, P-4710-057, Braga, Portugal
[7]Momentive Performance Materials, 22557 West Lunn Road, Strongsville, OH 44070, USA

*We report a bipolar field effect tunneling transistor that exploits to advantage the low density of states in graphene and its one atomic layer thickness. Our proof-of-concept devices are graphene heterostructures with atomically thin boron nitride acting as a tunnel barrier. They exhibit room temperature switching ratios ≈50, a value that can be enhanced further by optimizing the device structure. These devices have potential for high frequency operation and large scale integration.*

The performance of graphene-based field effect transistors (FETs) has been hampered by graphene's metallic conductivity at the neutrality point (NP) and the unimpeded electron transport through potential barriers due to Klein tunneling, which limit the achievable ON-OFF switching ratios to $\sim 10^3$ and those achieved so far at room temperature to <10 [1-7]. These low ratios are sufficient for individual high-frequency transistors and analogue electronics [4-7] but they present a fundamental problem for any realistic prospect of graphene-based integrated circuits [1-7]. A possible solution is to open a band gap in graphene, for example by using bilayer graphene [8,9], nanoribbons [10,11], quantum dots [11] or chemical derivatives [12] but it has proven difficult to achieve high ON-OFF ratios without degrading graphene's electronic quality.

In this report, we demonstrate an alternative graphene transistor architecture, namely a field-effect transistor based on quantum tunneling [13-17] from a graphene electrode through a thin insulating barrier (in our case, boron nitride of a nm thickness). The operation of the device relies on the voltage tunability of the tunneling density of states (DoS) in graphene and of the effective height $\Delta$ of the tunnel barrier adjacent to the graphene electrode.

The structure and operational principle of our FET are shown in Fig. 1. For convenience of characterization, both source and drain electrodes were made from graphene layers in the multiterminal Hall bar geometry [18]. This allows us to measure not only the tunnel current-voltage curves (I-V) but also the behavior of the graphene electrodes, thus providing additional information about the transistor operation. The tunnel barrier is hexagonal boron-nitride (hBN), and the core graphene-hBN-graphene structure is encapsulated in hBN to allow higher quality of the graphene electrodes [19,20]. The whole sandwich is placed on top of an oxidized Si wafer that acts as a gate electrode (Fig. 1A,B). When a gate voltage $V_g$ is applied between the Si substrate and the bottom graphene layer ($Gr_B$), the carrier concentrations $n_B$ and $n_T$ in both bottom and top electrodes increase due to the weak screening by monolayer graphene [21], as shown schematically in Fig. 1C. The increase of the Fermi energy $E_F$ in the graphene layers leads to a reduction in $\Delta$ for electrons tunneling predominantly at this energy [18]. In addition, as shown in the figure, the effective height also decreases relative to the NP because the electric field penetrating through $Gr_B$ alters the shape of the barrier [22,23]. Furthermore, the increase in the tunneling DoS as $E_F$ moves away from the NP [21] leads to an increase in the tunnel current $I$. Depending on parameters, any of the above three contributions can dominate changes in $I$ with varying $V_g$. We emphasize that the use of graphene in this device architecture is critical because this exploits graphene's low DoS which, for a given change in $V_g$, leads to a much larger increase in $E_F$ compared to a conventional two-dimensional gas with parabolic dispersion (cf. [13-17]). This translates into much larger changes of both $\Delta$ and tunneling DoS.

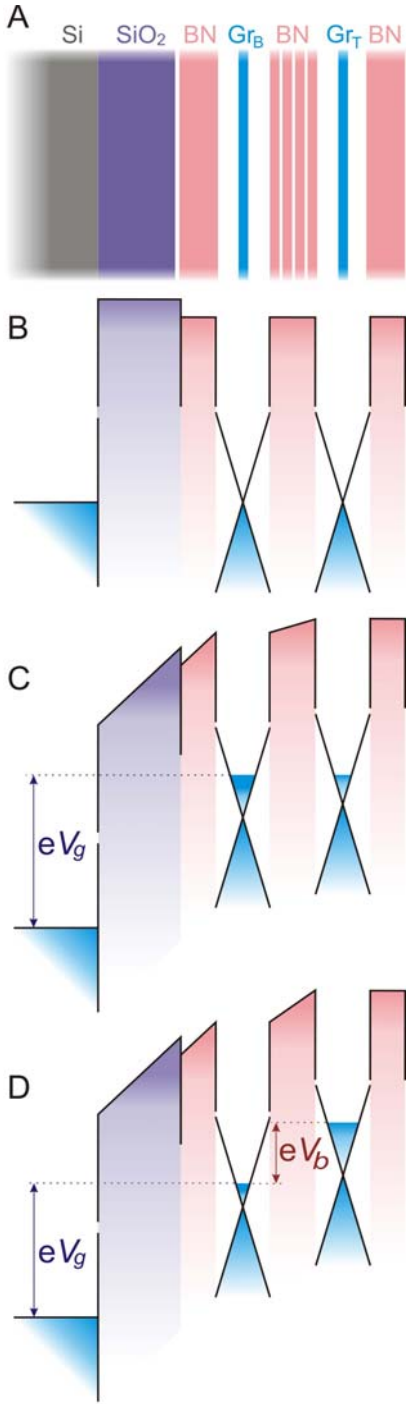

Fig. 1. Graphene field-effect tunneling transistor. (A) Schematic structure of our experimental devices. In the most basic version of the FET, only one graphene electrode ($Gr_B$) is essential and the outside electrode can be made from a metal. (B) The corresponding band structure with no gate voltage applied. (C) The same band structure for a finite gate voltage $V_g$ and zero bias $V_b$. (D) Both $V_g$ and $V_b$ are finite. The cones illustrate graphene's Dirac-like spectrum and, for simplicity, we consider the tunnel barrier for electrons.

To fabricate the device shown in Fig. 1A, we first prepared relatively thick hBN crystals on top of an oxidized Si wafer (300 nm of $SiO_2$) using the standard cleavage technique [24]. The crystals served as a high-quality atomically-flat substrate [19]. Monolayer graphene ($Gr_B$) was then transferred onto a selected hBN crystal (20-50 nm thick) using a dry transfer procedure [19,25]. After deposition of metal contacts (5nmTi/50nm Au) and etching to form a multiterminal Hall bar mesa, the structure was annealed at 350°C in hydrogen-argon atmosphere. A few-atom-thick hBN crystal was identified [26] and transferred on top of $Gr_B$ by using the same procedures. This hBN layer served as the tunnel barrier. The whole process of positioning, annealing and defining a Hall bar was repeated to make the second (top) graphene electrode ($Gr_T$). Finally, a thick hBN crystal encapsulated the entire multilayer structure (Fig. 1A; also, see [18]). Further details of our multistep fabrication procedures can be found in refs. [18,25]. We tested devices with tunnel barriers having thickness $d$ from 1 to 30 hBN layers [18]. To illustrate the basic principle of the tunneling FETs, we focus on the data obtained from four devices with a tunnel barrier made of 4-7 layers and discuss the changes observed for other $d$.

Fig. 2A shows the behavior of in-plane resistivity $\rho$ for the $Gr_B$ and $Gr_T$ layers as a function of $V_g$. The curves indicate little residual doping for encapsulated graphene ($\approx 0$ and $<10^{11}$ cm$^{-2}$ for $Gr_B$ and $Gr_T$, respectively). In both layers, $\rho$ strongly depends on $V_g$ showing that $Gr_B$ does not screen out the electric field induced by the Si-gate electrode. The screening efficiency can be quantified by Hall effect measurements (Fig. 2B-D). They show that the gate induces approximately the same amount of charge in both layers at low concentrations; that is, there is little screening if $n_B$ is small. As the concentration in $Gr_B$ increases, the $n_B(V_g)$ and $n_T(V_g)$ dependences become super- and sub-linear, respectively (Fig. 2B-C). This is due to the increase in $n_B$ which leads to an increasingly larger fraction of the gate-induced electric field being screened out by $Gr_B$ [18]. Hence more electrons accumulate in the bottom graphene electrode and fewer reach the top electrode. The total charge accumulated in both layers is linear in $V_g$

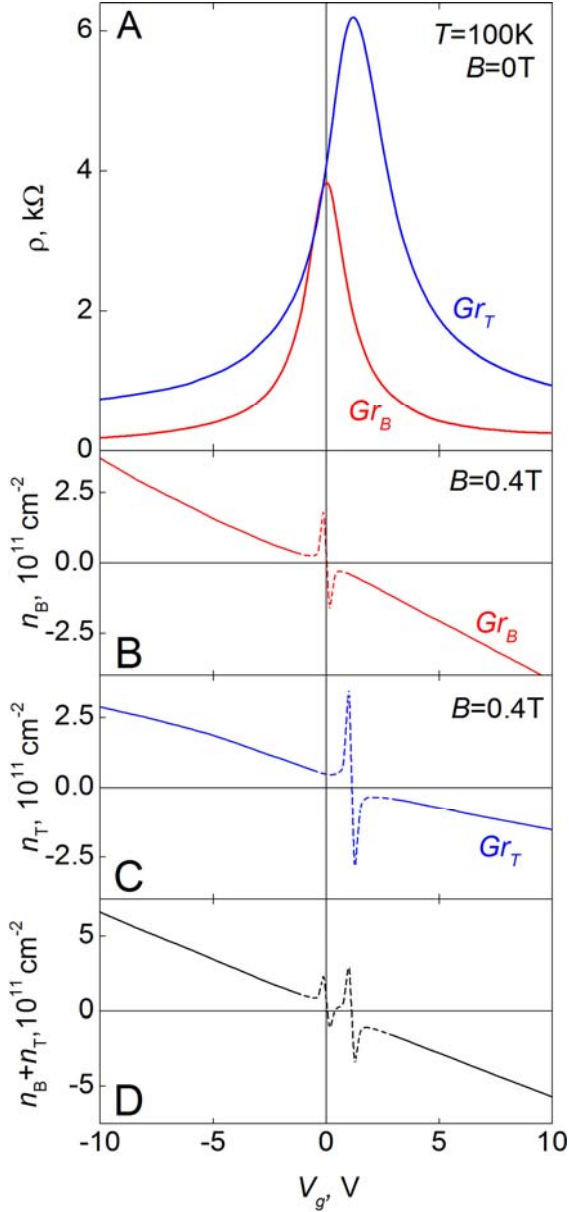

Fig. 2. Graphene as a tunneling electrode. (A) Resistivities of the source and drain graphene layers as a function of $V_g$. (B-D) Carrier concentrations in the two layers induced by gate voltage, which were calculated from the measured Hall resistivities $\rho_{xy}$ using the standard expression $n = B/e\rho_{xy}$ where $B$ is the magnetic field and $e$ the electron charge. Close to the NP, the spikes appear (shown by dotted curves) because the above expression is not valid in the inhomogeneous regime of electron-hole puddles. The shown device has a 4-layer hBN barrier. Carrier mobilities in the top and bottom layers are 35,000 and 90,000 cm$^2$/Vs,

(Fig. 2D), as expected. We can describe the observed redistribution of the charge between the two graphene layers in terms of the corresponding sequential circuit including the quantum capacitance [13,27] of the graphene layers (see [18]). Note that, for a parabolic band, the ratio between $n_B$ and $n_t$ would be independent on $V_g$ and, therefore, the electric field penetrating into the tunnel barrier would be significantly reduced even in the limit of zero $n_B$ [13,18].

A bias voltage $V_b$ applied between Gr$_B$ and Gr$_T$ gives rise to a tunnel current through the thin hBN barrier which scales with device area. Fig. 3A shows I–V characteristics for one of our devices at various $V_g$. First, we consider the case of zero $V_g$. At low $V_b$, $I$ is linear in bias, yielding tunnel resistivity $\rho^T = V_b/I \approx 100$ GΩ·μm$^2$ for this hBN thickness. At higher voltages ($V_b$ above ~0.1V), $I$ grows more rapidly. The I-V curves can be described (inset in Fig. 3A; also, see [18]) by the standard quantum-tunneling formulae [22,23] assuming energy conservation but no momentum conservation at the mismatched graphene-hBN interface [28]. As shown below, we can distinguish experimentally between electron and hole tunneling and find that the tunneling is due to holes. This is in agreement with a recent theory for the graphene-hBN interface [29], which reports a separation between the Dirac point in graphene and the top of the hBN valance band of ≈1.5eV whereas the conduction band is >4eV away from the Dirac point. The fit to our data with $\Delta = 1.5$eV yields a tunneling mass $m \approx 0.5$ m$_0$ (m$_0$ is the free electron mass), in agreement with the effective mass for holes in hBN [30]. Furthermore, our analysis indicates that $I$ varies mainly due to the change in the tunneling DoS, whereas the change in tunneling probability with applied bias is a significant but secondary effect [18]. This is due to the fact that, for our atomically-thin barriers with relatively low $\rho^T$, we are not in a regime of exponential sensitivity to changes in $\Delta[E_F(V_b)]$.

To demonstrate the transistor operation, Fig. 3A plots the influence of gate voltage on $I$. $V_g$ significantly enhances the tunnel current and the changes are strongest at low bias. The field effect is rather gradual for all gate voltages up to ±50V, a limit set by the electrical breakdown of our SiO$_2$ gate dielectric at typically ≈60V. To quantify this behavior, Fig. 3B plots the low-bias tunneling conductivity $\sigma^T = I/V_b$ as a function of $V_g$. The influence of $V_g$ is clearly asymmetric: $\sigma^T$ changes by a factor of $\approx 20$ for negative $V_g$ (holes) and by a factor of 6 for positive $V_g$ (electrons). We observed changes up to $\approx 50$ for hole tunneling in other devices and always the same asymmetry [18]. Also, the I-V curves of the devices showed little change between room and liquid-helium temperatures, as expected for $\Delta \gg$ thermal energy.

To analyze the observed behavior of $\sigma^T(V_g)$, we modeled the zero-bias conductivity by using the relation $\sigma^T \propto \mathrm{DoS}_B(V_g) \times \mathrm{DoS}_T(V_g) \times T(V_g)$, where the indices refer to the two graphene layers and $T(V_g)$ is the transmission coefficient through the hBN barrier [22,23]. The resulting curve shown in Fig. 3B explains qualitatively the main features in the measured data, using self-consistently the same tunneling parameters $m$ and $\Delta$ given above. At $V_g$ close to zero, corresponding to tunneling from states close to the NP, the tunneling DoS in both graphene layers is small and non-zero, due to residual doping, disorder and temperature [18]. The application of a gate voltage of either polarity leads to a higher DoS and, therefore, higher $\sigma^T$. The gradual increase in $\sigma^T(V_g)$ for both polarities in Fig. 3B is therefore due to the increasing DoS. However, $V_g$ also affects the transmission coefficient. Due to the shift of $E_F$ with changing $V_g$, the effective barrier height $\Delta$ decreases for one sign of charge carriers and increase for the other (Fig. 1B). This explains the asymmetry in both experimental and calculated $\sigma^T(V_g)$ in Fig. 3B: It is due to the change in $T(V_g)$. This clearly shows that for our devices the effect of $V_g$ on $T(V_g)$ is relatively weak (non-exponential) and comparable with the effect due to changes in the tunneling DoS. The sign of the asymmetry infers that the hBN barrier height is lower for holes than for electrons, in agreement with the graphene-hBN band structure calculations [29]. The weaker dependence of $I$ on $V_g$ at high bias is also understood in terms of the more gradual increase in the tunneling DoS and in $E_F$ at high doping ($V_b$ =0.5V correspond to $n_B \approx 10^{13}$ cm$^{-2}$).

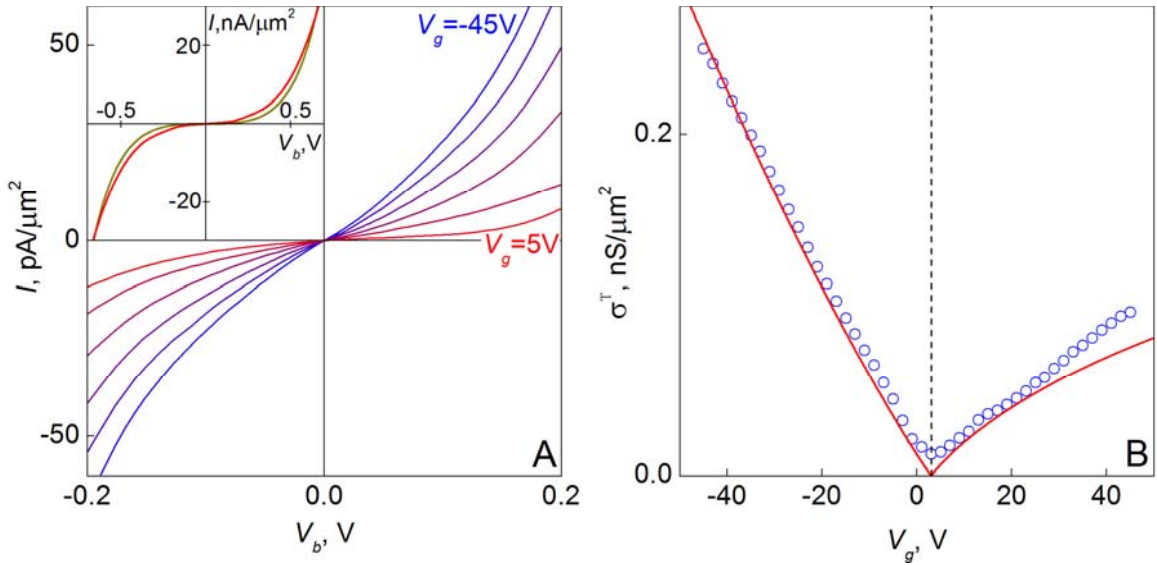

Fig. 3. Tunneling characteristics for a graphene-hBN device with 6±1 layers of hBN as the tunnel barrier. (A) I-Vs for different $V_g$ (in 10V steps). Note, that due to finite doping, the minimum tunneling conductivity is achieved at $V_g \approx 3$V. The inset compares the experimental I-V at $V_g$=5V (red curve) with theory (dark) which takes into account the linear DoS in the two graphene layers and assumes no momentum conservation. Further examples of experimental curves and their fitting can be found in Supplementary Material. (B) Zero-bias conductivity as a function of $V_g$. The symbols are experimental data, and the solid curve is our modeling. The curve is slightly shifted with respect to zero $V_g$ because of remnant chemical doping. In all the calculations, we assumed the hole tunneling with $m$ =0.5$m_0$ and $\Delta \approx 1.5$ eV [29,30] and used $d$ as measured by atomic force microscopy. Both $I$ and $\sigma$ are normalized per tunnel area, which was typically 10 to 100 μm$^2$ for the studied devices. Temperature: 240 K.

Our results and analysis suggest that higher ON-OFF ratios could be achieved by using either higher $V_g$ or making devices with larger $d$, so that the tunneling depends exponentially on bias and is controlled by the barrier height rather than the DoS. The former route is limited by the electrical breakdown of dielectrics at ~1V/nm ($V_g \approx$ 300V for our SiO$_2$ thickness). By extrapolating the analysis shown in Fig. 3B to such voltages, we find that ON-OFF ratios >$10^4$ would be possible for our 4-7 layer devices if SiO$_2$ of highest quality were used. However, it would still require unrealistically large $V_g$ to enter the regime where $E_F$ becomes comparable with $\Delta$ and changes in $\sigma^T(V_g)$ are exponentially fast. Therefore, we have tried the alternative option and investigated devices with both thinner and thicker hBN barriers. For 1 to 3 hBN layers, we find that zero-bias $\sigma^T$ increases exponentially with decreasing number of layers, consistent with quantum tunneling, and we observe a weaker influence of $V_g$ on $I$, as expected for the more conductive regime. On the other hand, the thicker hBN barriers are prone to electrical breakdown. Nonetheless, for a few devices with $d \approx$ 6 to 9 nm we were able to measure a tunnel current without breakdown. A significant current (>10pA) appeared at biases of several volts and increased exponentially with $V_b$. The thicker devices' I-V characteristics could be fitted using the same hole-tunneling parameters used above, thus indicating quantum tunneling rather than an onset of electrical breakdown. Unfortunately, no significant changes (exceeding 50%) in the tunnel current could be induced by $V_g$. This insensitivity to gate voltage remains to be understood but, partially, is due to high $n_B$ (>$10^{13}$ cm$^{-2}$) induced by $V_b$, so that Gr$_B$ becomes strongly metallic and efficiently screens the influence of the gate.

We conclude that our tunneling devices offer a viable route for high speed graphene-based analogue electronics. The ON-OFF ratios already exceed those demonstrated for planar graphene FETs at room temperature by a factor of 10 [3-7]. The transit time for the tunneling electrons through the nm-thick barriers is expected to be extremely fast (a few fs) [13-17] and exceeds the electron transit time in submicron planar FETs. It should also be possible to decrease the lateral size of the tunneling FETs down to the 10 nm scale, a requirement for integrated circuits. Furthermore, there appears to be no fundamental limitation to significantly enhance the ON-OFF ratios by optimizing the architecture and by using higher-quality dielectrics. We believe that the electronic properties of the demonstrated devices merit further research to explore their limitations and scope, and their potential for applications.

**Supplementary Material**

### #1 *Experimental structures*
Our devices contain two graphene Hall bars placed on top of each other with a thin layer of hBN in between. Figure S1 shows one of the studied devices. The turquoise area in Fig. S1A is a thick hBN crystal on top of an oxidized Si wafer (brown-purple). The hBN layer served as a substrate to ensure the quality of the bottom graphene electrode. The actual graphene-hBN-graphene-hBN sandwich is highly transparent and practically invisible on this image taken in an optical microscope (Fig. S1A). Nonetheless, one may discern a mesa structure in the central area between the Au leads. The multilayer Hall bar geometry is illustrated in Fig. S1B. This is an electron micrograph of the same device but before depositing Au contacts. The colored image of various layers was used at a design stage for the last round of electron-beam lithography. The Au leads (deposited later) are shown in violet, and two graphene mesas in orange and green. The hBN crystal used as the tunnel barrier can be seen as a light grey patch of irregular shape. Its thickness was determined using atomic force microscopy, Raman microscopy and optical contrast [26].

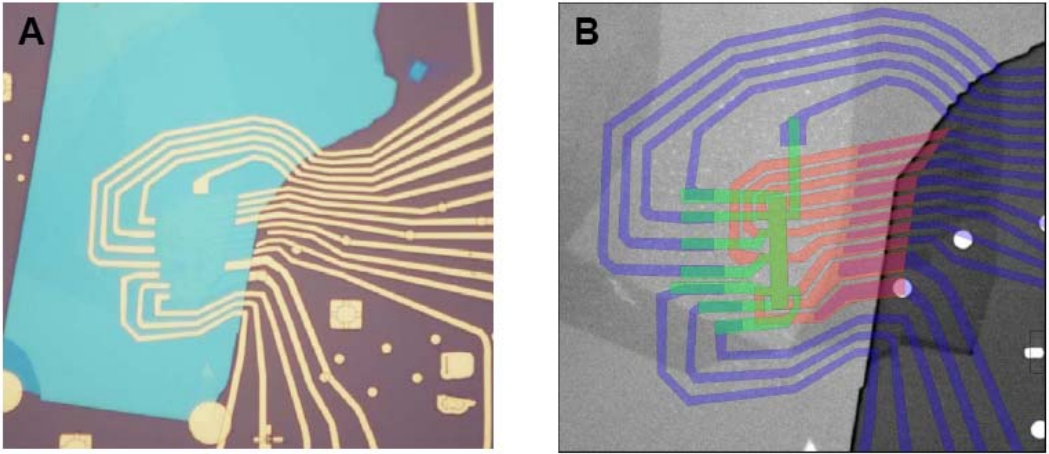

Figure S1. One of our hBN-graphene-hBN-graphene-hBN devices. (A) Optical image of the final device. (B) Electron micrograph of the same device at the final design stage before evaporating Au leads. Two 10-terminal Hall bars made from graphene are shown in green and orange. The spatial scale is given by the width of the Hall bar, which was 2 μm for this device. Fabrication required 4 dry transfers and alignments of the graphene and hBN crystals, 4 nonconsecutive rounds of electron-beam lithography, 3 rounds of plasma etching and two separate metal depositions.

### #2 *Penetration of electric field through the graphene electrode*
Consider the geometry shown in Fig. 1A of the main text. The external electric field between the Si and bottom graphene electrodes, which are separated by distance $D$, is $F_g = V_g/D$ (dielectric constants for both $SiO_2$ and hBN are similar and, for simplicity, we assume them both equal to $\varepsilon$). The electric field $F_b$ between $Gr_B$ and $Gr_T$ and the induced carrier densities in the graphene plates $n_T$ and $n_B$ are related by the equations

$$\varepsilon(F_b - F_g) = 4\pi n_B e$$
$$-\varepsilon F_b = 4\pi n_T e$$

A bias voltage $V_b$ between the two graphene electrodes is given by

$$eV_b = eF_b d - \mu(n_T) + \mu(n_B)$$

where $d$ is the hBN thickness and $\mu(n)$ are the chemical potentials in the corresponding graphene layers. For simplicity, we assume that graphene electrodes are chemically undoped and, therefore, $n_T = n_B = 0$ in the absence of applied voltages.

Taking into account the electron-hole symmetry $\mu(-n) = -\mu(n)$, we obtain the following equation

$$\frac{4\pi e^2 d}{\varepsilon} n_T + \mu(n_T) + \mu\left(n_T + \frac{\varepsilon F_g}{4\pi e}\right) + eV_b = 0 \tag{S1}$$

which allows us to determine $n_T$ induced by the field effect in $Gr_T$ for a given $V_g$. For a conventional two-dimensional (2D) electron gas, $\mu(n) \propto n$ and the first term in eq. (S1), which describes the classical capacitance of the tunnel barrier, is dominant for any realistic $d$, larger than interatomic distances. In graphene with its low DoS and Dirac-like spectrum, $\mu(n) \propto \sqrt{n}$ and this leads to a qualitatively different behavior, which can be described in terms of quantum capacitance [27] (also note the discussion of doping of graphene through an hBN spacer in ref. [S1]).

The above expressions were employed to find $n_T$ and $n_B$ as a function of bias $V_b$ and gate voltage $V_g$ and the results were then used to model the I-V characteristics (see the theory curves in Fig. 3 of the main text). To illustrate the agreement between the experiment and theory at the intermediate stage of determining $n_T$ and $n_B$, Figure S2 shows the same experimental data for carrier concentrations in the top and bottom graphene layer $n(V_g)$ as in Fig. 2B,C and compares them with the behavior expected from solving eq. (S1).

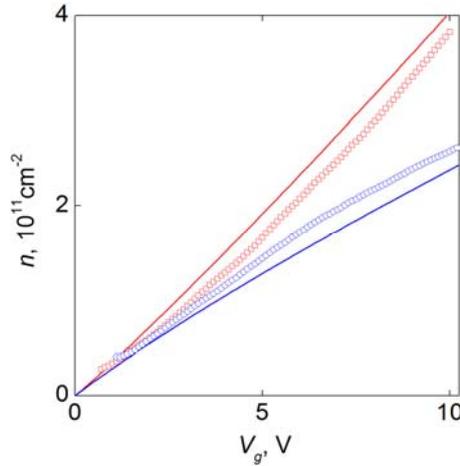

Figure S2. Nonlinear dependence of charge carrier concentrations in the two graphene electrodes as a function of gate voltage. The symbols are experimental data (red symbols for the bottom graphene layer; blue for the top). The solid curves in the corresponding colors are our modeling. No fitting parameters are used.

#3 *Modeling of device operation*
I-V curves for a tunnel junction are generally described by [23]

$$I(V) \propto \int dE\, DoS_B(E)\, DoS_T(E-eV)\, T(E)[f(E-eV) - f(E)] \tag{S2}$$

where $f(E)$ is the Fermi distribution function. At low temperatures the difference of the Fermi functions restricts the relevant energy $E$ integral to $\mu < E < \mu + eV$ where $\mu$ is the chemical potential and, to be specific, we consider the case $eV > 0$. The above formula assumes that there is no in-plane momentum conservation, which is most likely to be the case of realistic graphene-hBN interfaces. Indeed, there are several possible mechanisms for elastic scattering at the interface and, in particular, unavoidable fluctuations of the mass term due to the lattice mismatch [S2]. Note that elastic tunneling is forbidden between two 2D systems if in-plane momentum is conserved.

If the tunneling conductance per channel is much smaller than conductivity quantum $e^2/h$ (as in our case) the transmission probability $T$ is exponentially small and depends strongly on the energy $E$ of tunneling electrons,

$$T(E) = A(E)\exp[-W(E)] \tag{S3}$$

where *A* is a smooth function that depends on details of the wave-function matching at the interface. In our modeling, we assume *A=const.*

Let us now discuss common functional forms for *W(E)*. For the case of an isotropic barrier, we need to solve the dispersion equation $E = \varepsilon_n(k_x, k_y, k_z)$ for each band of the barrier material, where *E* is the energy of electrons tunneling in the *z* direction. No real solution for $k_z$ is possible inside the energy gap, and the minimal Im$k_z$ for a given *E* and arbitrary $k_x$ and $k_y$, which dominates the tunneling probability, is given by

$$W(E) = 2d \, \mathrm{Im} \, k_z$$

For the case of parabolic bands, $\mathrm{Im} \, k_z = \frac{\sqrt{2m\Delta}}{\hbar}$ where Δ is the barrier height (in our case, the distance to the valence band) and *m* is the effective mass [22-23,S3].

In the case of layered crystals, their band structure can be described in the simplest approximation as

$$\varepsilon(k_x, k_y, k_z) = \tau(k_z) + \varepsilon_1(k_x, k_y) \tag{S4}$$

where $\tau(k_z) = 2t_\perp \cos(k_z l)$; $t_\perp$ describes the interlayer coupling and *l* is the interlayer distance (for the case of hBN, $l \approx 3.4$Å). By solving the corresponding tunneling equation, we find $k_z$ within the gap to be

$$k_z = \frac{i}{l} \ln\left( \left| \frac{E - \varepsilon_1}{2t_\perp} \right| + \sqrt{\left(\frac{E - \varepsilon_1}{2t_\perp}\right)^2 - 1} \right)$$

The top of the valence band corresponds to $E_{\max} = \max \varepsilon_1(k_x, k_y) + 2t_\perp$ (to be specific, we choose $t_\perp > 0$), and the optimal value for the tunneling wavevector is then

$$\mathrm{Im} \, k_z = \frac{1}{l} \ln\left( \left(\frac{\Delta}{2t_\perp} + 1\right) + \sqrt{\left(\frac{\Delta}{2t_\perp} + 1\right)^2 - 1} \right) \tag{S5}$$

where $\Delta = E - E_{\max}$. If $\Delta \gg 2t_\perp$, this expression can be simplified as $k_z = \frac{i}{l} \ln\left(\frac{\Delta}{t_\perp}\right)$ and yields the tunneling probability $T \propto (t_\perp/\Delta)^{2n}$ where $n = d/l$ is the number of atomic layers in the tunnel barrier. In the opposite limit of $\Delta \ll 2t_\perp$, we obtain $k_z = \frac{i}{l}\sqrt{\frac{\Delta}{t_\perp}} = i\frac{\sqrt{2m^*\Delta}}{\hbar}$ where $m^* = \frac{\hbar^2}{2t_\perp l^2}$ is the effective mass in the tunneling direction. This shows that the standard isotropic model is applicable to layered crystals, provided tunneling occurs not too far from the band-gap edge.

Eq. (S4) is a simplified version of the real band structure of hBN, which depends on stacking order. hBN crystals usually have AA' stacking [S4]. In the next approximation that allows an analytical solution by neglecting the mixing of π and σ bands [29,30], we obtain the following dispersion relation [S4]

$$\varepsilon^2(k_x, k_y, k_z) = \frac{E_g^2}{4} + \tau^2(k_z) + \varepsilon_1^2(k_x, k_y) \pm 2\tau(k_z)\varepsilon_1(k_x, k_y) \tag{S6}$$

where $E_g$ is the energy difference between boron and nitrogen sites [S4]. In this case, we find

$$\mathrm{Im} \, k_z = \frac{1}{l} \ln\left( \frac{\Phi}{2t_\perp} + \sqrt{\left(\frac{\Phi}{2t_\perp}\right)^2 - 1} \right) \tag{S7}$$

where $\Phi = \sqrt{E^2 - \frac{E_g^2}{4}} - |\varepsilon_1(k_x, k_y)|$. Eq. (S7) differs from (S5) by replacement $E \to \sqrt{E^2 - E_g^2/4}$, which indicates the general validity of equation Im$k_z \propto \ln(\Delta)$ for describing vertical tunneling through strongly layered materials. (S5) and (S7) fit our experimental data equally well. It is worth noting that the tunneling exponent through layered crystals depends on *E* only weakly (logarithmically) in comparison with isotopic crystals that exhibit the standard square-root energy dependence. For small changes in $\Delta$, this difference is unimportant (see below).

Finally, in the case of a strong electric field such that it changes the rectangular shape of the tunnel barrier (Fig. 1D), the above expressions for $W$ can be generalized within the WKB approximation [S3] as

$$W = 2\int_0^d dx\, \mathrm{Im}\, k_z(\Delta \to \Delta(x)).$$

#4 *Layered versus isotropic barrier*
In the main text, we have chosen for the sake of brevity to ignore the fact that our tunnel barriers are made from a strongly layered material. This simplification allowed us to refer to the standard tunneling theory. However, the assumption can be justified further by the fact that, for our device parameters, we have found no difference between the I-V characteristics calculated for the layered and isotropic materials and, therefore, we cannot distinguish between the two cases. To illustrate the indifference to the layered structure of our tunnel barrier, Figure S3 shows experimental I-V curves for two devices and compares them with the behavior expected for layered and isotropic cases. No major difference can be seen, except at low bias in Fig. S3A. The exact shape of experimental curves at low bias varies from sample to sample (cf. Fig. S3A & B) and, hence, we do not discuss the difference.

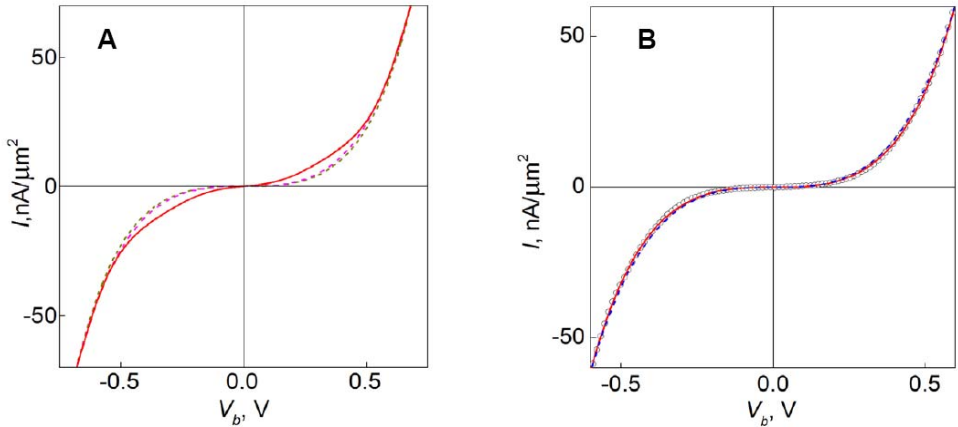

Figure S3. Tunneling I-V characteristics for two different 4-hBN-layer devices at zero gate voltage and their comparison with theory. (A) The red solid curve is the experimental data from Fig. 3. The two dashed curves are our modeling for an isotropic barrier ($\Delta$ and $m$ as in the main text) and for a layered barrier of the same height and $t_\perp$ =0.6eV, by using formulae from the above section. Note that $t_\perp$ ≈0.6eV corresponds to $m$ =0.5$m_0$. (B) Nominally similar device (for clarity, the experimental data are shown by symbols). The curves are again the layered and isotropic versions of the tunneling theory. The fitting parameter is the constant $A$ in eq. (S3), which determines the absolute value of $I$. The close agreement between functional forms of the theoretical curves validates the use of the conventional tunneling formulae in the main text.

#5 *Additional examples of our device operation*
We have studied 6 multiterminal devices such as shown in Fig. S1 and >10 simpler tunneling FETs with only one or two Ohmic contacts attached to each graphene electrode. The latter type does not provide much information about the properties of the graphene electrodes but even one contact is sufficient to study their tunneling I-V characteristics. The devices with the same hBN thickness have exhibited qualitatively similar behavior, as discussed in the main text. To illustrate the degree of reproducibility for different samples, Figure S4 plots the behavior observed in another device with the tunnel barrier consisting of 4 hBN layers. One can see that the nonlinear I-V characteristics are qualitatively similar to those presented in the main text, and their response to gate voltage is also fairly similar.

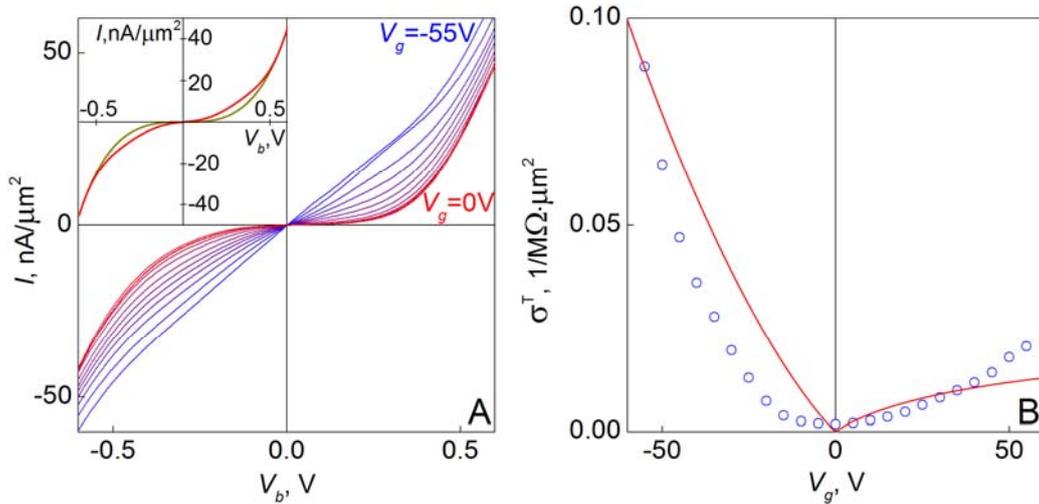

Figure S4. Another hBN-graphene-hBN-graphene-hBN field-effect device. (A) Tunneling I-Vs and their response to gate voltage (in 5V steps, cf. Fig. 3 of the main text). The inset compares the experimental I-V at zero gate voltage (red curve) with theory (dark) which takes into account the linear DoS in the two graphene layers and assumes no momentum conservation. Temperature: 300 K. (B) Changes in low-bias tunneling (symbols) and the theory fit for 4 hBN layers (solid curve). The main difference with respect to the device in the main text is a weak response at low gate voltages, which is probably due to stronger disorder and chemical doping that smears the gate influence. The electron-hole asymmetry again implies the hole tunneling as discussed in the main text.

The only consistent difference that we noticed for a number of devices with 4 or more atomic layers of hBN was the absolute value of $\sigma^T$ which could vary by a factor of 100 for nominally the same *d*. Although this can be attributed to possible errors in determining the number of layers in thicker hBN [26], more careful analysis of the devices' response to bias and gate voltages reveals that the reason for these variations is more likely to be inhomogeneous thickness of hBN. We believe that in some devices one or two layers can be missing locally (in submicron scale patches) so that the tunnel current then concentrates within these thinner areas. Graphite is known to cleave leaving occasional stripes of smaller thickness for few-layer graphene crystals and, whereas it is possible to see missing graphene patches in an optical microscope, hBN does not allow the required resolution [26].